# Energy transport by classical waves through multilayers of diffusing slabs

**Sijmen Gerritsen, and Gerrit E.W. Bauer**
*Kavli Institute of Nanoscience, Delft University of Technology, Lorentzweg 1, 2628CJ Delft, The Netherlands*
*s.gerritsen@tnw.tudelft.nl*

**Abstract:** We describe the effect of interfaces on classical wave propagation through diffusing layered media. A series resistor model for wave energy transport is introduced and we derive a microscopic expression for the interface resistance.
©2006 Optical Society of America
**OCIS codes:** (290.1990) Diffusion; (290.4210) Multiple scattering

## 1. Introduction

Multiple scattering at small inhomogeneities in disordered (complex) media often limits the applicability of imaging and detection techniques that make use of classical waves, like electromagnetic waves in radar applications and optical tomography [1], or acoustic waves in ultrasonical imaging and geophysical exploration [2]. When scattering is so weak that a substantial ballistic signal remains, larger objects with sufficient contrast in the constitutive parameters can be imaged by measuring travel times of pulsed sources. When the travel path of the waves exceeds the mean free path, wave energy is propagated diffusively, which complicates imaging and detection considerably [3]. A formalism that relates diffusive wave propagation to the material properties in such systems in a simple physical picture could be useful for applications. However, describing systems that contain both many small scatterers (on the wavelength scale) and macroscopic objects using the diffusion equation requires careful consideration of the boundary conditions at the interfaces [4].

In this paper, we focus on the influence of (sharp) interfaces between diffusing slabs on diffusive wave propagation. When a diffusive energy current flows through multilayers of slabs of disordered materials in the steady state, a gradient in diffuse energy density is built up. Across (sharp) interfaces between different slabs, the diffuse energy density drops in discontinuous steps. We derive a formalism that relates this drop to a potential difference by introducing an interface resistance. We show that the similarities between classical diffusive wave propagation and electronic/phonon transport can be used to develop a series resistor model analogous to the description of electric current in metallic multilayers. Besides being a useful alternative description for wave propagation in diffuse systems, it is an example of the application of ideas originally developed to describe electron transport in mesoscopic physics to classical wave propagation in complex media.

## 2. Wave propagation between reservoirs through leads

Let us first consider energy transport by classical waves between two reservoirs (or cavities) through leads in the language of scattering theory of transport pioneered by Landauer and Büttiker [5]. We focus on a ballistic constriction schematically represented in Fig. 1. The reservoirs are in local equilibrium and kept at a constant energy density $W_i$ where $i=L(R)$ for the left(right) hand side, which are different with respect to the material properties. Energy is equipartitioned over the modes in the reservoirs so that every mode with frequency $\omega$ carries the same energy, determined by a distribution function $f_i$. For simplicity, we only consider a single polarization (acoustic waves in liquids). The reservoirs are assumed to be adiabatically coupled to ideal leads and the energy current can be expressed in terms of the difference in energy distribution in the reservoirs by [6],

$$J = \frac{1}{2\pi}\int_0^\infty d\omega\, \hbar\omega \left[ f_L(\omega) - f_R(\omega) \right] \sum_\alpha T_\alpha(\omega). \tag{1}$$

The summation indexed by $\alpha$ goes over the (quasi one-dimensional) right(left) going modes in the left(right) lead and $T_\alpha$ is the probability that the mode is transmitted into the other lead. When the leads and reservoirs left and right are identical, $T_\alpha=1$ and the current is just proportional to the number of modes in the leads at a certain frequency, $N(\omega)$.

When the wave velocity (in the continuum limit) between the left and right leads and reservoirs is different, the transport is affected by the reflection at the interface. We consider the case when the leads are wide (compared to the wavelength) so there are many propagating modes. In this case, we may index the modes with the parallel (to the interface) wave vector $\mathbf{k}_\parallel$, which relates the modes to the wave vector $\mathbf{k}$ of the (plane wave) modes in the reservoirs. When we sum over the modes in the left lead, $k = \omega/c_L$, where $c_L$ is the wave velocity on the left hand side. The summation over modes can then be expressed as

$$\sum_\alpha T_\alpha(\omega) = \sum_{\mathbf{k}_\parallel}^{k_\parallel \leq k} T_{\mathbf{k}_\parallel}(\omega) = N_L(\omega)\overline{T}_{LR}(\omega), \qquad (2)$$

where $\overline{T}_{LR}$ is the integrated total energy transmission and for a wide lead of cross-sectional area $A$, $N_L = A\omega^2/(4\pi c_L^2)$.

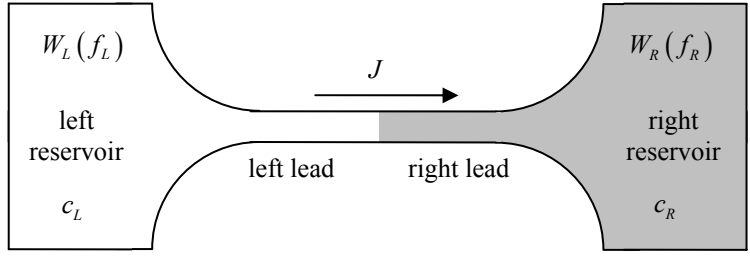

Fig. 1. Schematic representation of the reservoir-lead-lead-reservoir configuration.

The energy density in the reservoirs is given by

$$W_i = \frac{1}{2\pi^2}\int d\omega \hbar\omega \frac{\omega^2}{c_i^3} f_i(\omega). \qquad (3)$$

When the total energy transmission $\overline{T}_{LR}$ does not depend on frequency, or the energy distribution is strongly peaked at a certain frequency, we can relate the current to the difference in energy density and wave velocity by

$$J = \frac{A}{4}\frac{1}{c_L^2}\overline{T}_{LR}\left(c_L^3 W_L - c_R^3 W_R\right) = \frac{1}{R_{LR}}(\phi_L - \phi_R). \qquad (4)$$

So the energy current is not driven by the difference in the energy densities, but by the difference in a potential $\phi_i = c_i^3 W_i$. The resistance $R_{LR} = 4c_L^2/(A\overline{T}_{LR})$ can thus be expressed in terms of the microscopic scattering matrix of the interface. When $c_L = c_R = c$ only a geometrical construction remains with a finite resistance that in electronic transport is called the Sharvin resistance. For this we obtain from the number of modes in the leads that

$$R_{Sh} = \frac{4}{A}c^2. \qquad (5)$$

The total transmission can be computed readily for a "clean" specular interface between two homogeneous wide leads by integrating over the angle resolved (plane wave) transmission probabilities given by the well-known Fresnel coefficients. The final result depends only on the velocity contrast ($\gamma = c_R/c_L$) between the left and right leads:

$$\overline{T}_{LR} = \gamma\frac{4(2+\gamma)}{3(1+\gamma)^2} \text{ for } \gamma < 1, \quad \overline{T}_{LR} = \frac{1}{\gamma^2}\frac{4(1+2\gamma)}{3(1+\gamma)^2} \text{ for } \gamma > 1. \qquad (6)$$

### 3. Energy current through multilayers and the interface resistance

*3.1 The single interface*

Let us now turn to propagation through interfaces between diffusing media. In this case we are again able to describe wave energy propagation in much the same way as electronic transport [7,8]. When two 3D diffusing slabs are put into contact and a diffusive current (from a planar source) is running from left to right, a gradient in diffusive energy density is built up. The distribution functions inside the slabs now only differ from the distribution in the reservoirs from the previous section by a finite drift in the direction of the energy current. We can regard the diffusing slabs as reservoirs and calculate the current through the interface that is driven by the jump in the distribution functions left and right from the interface. When the interface is transparent, the resistance should vanish (there is no Sharvin resistance). It has been shown in Ref. 8 that we can take this into account by subtracting the average of the Sharvin resistances from $R_{LR}$ to obtain the resistance of the interface between the two layers [8]:

$$R_{Int} = R_{LR} - R_{Sh} = \frac{4}{A} c_L^2 \left[ \overline{T}_{LR}^{-1} - \frac{1}{2}\left(1 + \frac{c_R^2}{c_L^2}\right) \right]. \tag{7}$$

Here the total transmission coefficient $\overline{T}_{LR}$ still has to be calculated from the microscopic scattering matrix of the (not necessarily clean) interface.

*3.2 Multilayers of diffusing slabs*

It is now straightforward to introduce a series resister model for the energy current through multilayers of diffusing slabs. We assume a planar source in a 3D layered structure. The current between any two planes indexed by $L$ and $R$ in the multilayer structure is related to the driving potential by

$$J = \frac{1}{R_{Tot}}(\phi_L - \phi_R), \tag{8}$$

where $\phi_i = c_i^3 W_i$ and the total resistance $R_{Tot}$ is just the sum of bulk and interface resistances of the layers between $L$ and $R$, where $R_{Int}$ is given by Eq. (7). Within the bulk layer, energy density and current are related through Fick's law ($J/A = -D\partial_z W$) and the resistance of a bulk layer is thus $R_i = L_i c_i^3 / AD_i$, where $L_i$ is the layer thickness (much larger than the mean free path) and $D_i$ the diffusion constant.

## 4. Conclusions

We relate the diffusive current through multilayers of diffusing slabs to the local energy density with a series resistor model ("Ohm's law"). We have shown that the driving potential difference is not the difference in energy density but the difference in the potential $\phi_i = c_i^3 W_i$, where $c_i$ is the local wave velocity and $W_i$ the local energy density. We have derived an expression for the interface resistance that relates the potential drop across an interface between two diffusing layers to the total energy current through the interface in terms of the microscopic interface matrix. In this way, we manage to describe the relation between the diffusive current and the diffusive energy density and the material properties of layered diffusing media in a simple physical picture.